# A Maslow-Inspired Hierarchy of Engagement with AI Model

**Prof Madara Ogot**

University of Nairobi

https://0000-0003-4010-2434 | madaraogot@uonbi.ac.ke

**Abstract**

The rapid proliferation of artificial intelligence (AI) across industry, government, and education highlights the urgent need for robust frameworks to conceptualise and guide engagement. This paper introduces the Hierarchy of Engagement with AI (HE-AI) model, a novel maturity framework inspired by Maslow's hierarchy of needs. The model conceptualises AI adoption as a progression through eight levels, beginning with initial exposure and basic understanding and culminating in ecosystem collaboration and societal impact. Each level integrates technical, organisational, and ethical dimensions, emphasising that AI maturity is not only a matter of infrastructure and capability but also of trust, governance, and responsibility. Initial validation of the model using four diverse case studies (General Motors, the Government of Estonia, the University of Texas System, and the African Union AI Strategy) demonstrate the model's contextual flexibility across various sectors. The model provides scholars with a framework for analysing AI maturity and offers practitioners and policymakers a diagnostic and strategic planning tool to guide responsible and sustainable AI engagement. The proposed model demonstrates that AI maturity progression is multi-dimensional, requiring technological capability, ethical integrity, organisational resilience, and ecosystem collaboration.

# 1. Introduction

Artificial intelligence (AI) systems are increasingly pervasive in our lives as they continue to evolve rapidly with increasing capabilities. They find use in all sectors, including healthcare, finance, education, national security, and climate mitigation and adaptation (Triguero et al., 2023; NIST, 2024). Currently, few structured maturity models (MMs) exist that provide frameworks for capturing individual and institutional AI journeys, from experimentation to mastery and responsible societal contribution, as well as holistic approaches to assess AI readiness, capability, and ethical maturity. The proposed framework captures how individuals and organisations learn to use, deploy and manage AI.

The importance of MMs cannot be overstated. They provide a shared vocabulary for individuals, organisations and policymakers to determine their progression on a continuum of AI adoption (Sadiq et al., 2023; Butler et al., 2023) and identify gaps between capabilities and aspirations. The gaps could be technical (e.g., lack of infrastructure), cultural (e.g., lack of trust), or educational (e.g., lack of AI literacy or skills) (Mäntymäki et al., 2022; Dotan et al., 2024). MMs support organisational resource allocation and investment planning by providing clear development stages and readiness indicators, thereby reducing the risk of misallocation of funds or deployment of AI tools in environments that cannot safely and effectively utilise them (Weill et al., 2024; Reichl & Gruenbichler, 2024). They enable individuals or organisations to assess their current state and plan realistic, structured journeys towards higher capabilities. Newly introduced government regulatory frameworks, such as the EU's AI Act (2021) and the US NIST AI Risk Management Framework (NIST, 2024), require organisations to demonstrate compliance, risk awareness, bias mitigation, and transparency. MMs can provide tools to define and communicate how these standards are being met (Steinmetz et al., 2025). They help to democratise AI by recognising that people start their AI journeys at diverse points. For example, most people's first encounter with AI was through curiosity or casual experimentation with free tools such as ChatGPT and not from formal training (Sadiq et al., 2023; NIST, 2024). MMs incorporating these early interactions promote inclusion and encourage learning.

Most AI MMs, however, target organisations, emphasising strategy alignment, data readiness and process integration (Reichl & Gruenbichler, 2024; Weill et al., 2024). They ignore individual user journeys. Process-oriented MMs, such as De Silva and Alahakoon's (2021) *AI lifecycle framework* or NIST's (2024) *Taxonomy of AI use cases,* also do not address why users engage with AI at different levels, how their capabilities and skills evolve, or what enables them to move from one level to the next. Ethics and trust maturity models, such as the *Trust Calibration Maturity Model* (Steinmetz et al., 2025) and the *Hourglass Model of AI Governance* (Mäntymäki et al., 2022), assess AI system risk and accountability through external evaluative lenses rather than as outcomes of an individual's or organisation's advancing capability and intent.

The proposed psychology-grounded Hierarchy of Engagement with AI (HE-AI) model aims to address some of these limitations, ranging from the earliest forms of AI exposure to technical mastery, responsible deployment, and community contribution. The model addresses individual motivations, technical skills, infrastructure, and organisational integration. The eight-level framework, inspired by Maslow's hierarchy of human needs (Maslow, 1943; 1970), traces AI journeys (individual and organisation) from Initial Exposure and Curiosity to Societal and Global Integration. Core motivations, technical skills, task complexity, and infrastructure engagements define each level. Differing from existing models that frame AI maturity as a one-dimensional ladder or organisational benchmark, the HE-AI model places

AI engagement within human motivational contexts. A literature review of key contributions to the development of AI maturity models is presented, highlighting their contributions and gaps that the proposed HE-AI model aims to address.

## 2. Literature Review

Most of the literature on AI MMs is focused on organisational models that primarily cover five domains (technological readiness, data capability, organisational culture, and ethical readiness) (Sadiq et al., 2023), and mainly rely on descriptive, static taxonomies (Reichl & Gruenbichler, 2024). For example, the *Enterprise AI MM* (Weill et al., 2024) has four stages (experimentation, pilot, industrialisation, and strategic deployment) through which organisations strengthen their AI capabilities. PwC's (2020) *AI Capability Maturity Framework* has five stages (aware, active, operational, systemic, and transformational). It measures the extent to which AI has been integrated into an organisation's core operations and strategy. Their framework evaluates capabilities across several dimensions (leadership, strategy, data, technology, and talent). Deloitte's (2021) *AI Maturity Framework also has* five stages (starting, experimenting, formalising, operationalising, and transforming) that are supported by two overarching domains of AI Foundations (data, technology, and talent) and AI strategy (governance, risk, key performance indicators, and innovation culture). PwC's and Deloitte's models, designed for enterprise-scale operations typically in data-rich, high-capacity environments, aim to help organisations optimally utilise AI to achieve a competitive advantage. De Silva and Alahakoon (2021) proposed a 17-stage lifecycle model for AI development to guide the implementation processes. Their framework does not address why actors engage with AI at different stages, how their motivations evolve, or how technical maturity develops over time.

Other AI MMs target public sector organisations. For example, Noymanee et al. (2022) developed a five-level maturity model (new entry level, elementary entry level, operational level, proficiency level 1, proficiency level 2) across four domains (strategy, organisation, technology, and data). Their model seeks to guide public sector organisations in assessing their readiness for AI implementation and long-term integration into workflows. Hudaib et al.'s (2024) model has five maturity levels (initial, opportunistic, systematic, transformative, and innovative) that are mapped across six dimensions (strategy, data, technology, process, people, and governance). They can be used as a diagnostic and strategic planning tool to measure organisational progress, guide policy and capability development, and assess resources and alignment with the organisation's AI strategies. Zuiderwijk et al.'s (2024) AI MM assesses the readiness and sophistication of public sector organisations in their AI adoption journeys. Their six-level MM (ad hoc, foundational, developing, defined, integrated, and transformative) focuses on institutional capabilities and policy coherence and assessed across several dimensions (AI vision and strategy, data management, ethics and responsibility, and the organisational ecosystem).

By design, organisational models assess firm-level progress. They do not incorporate an individual's psychological or developmental pathways or early engagement with AI. The proposed HE-AI model's broader scope views AI maturity through both individual and organisational lenses, capturing their inter-twinned AI journeys. It complements organisational models by bridging the gap between organisational readiness and pathways to building individual and organisational AI literacy and innovation. Levels 0 (Initial Exposure and Curiosity) and 1 (Awareness and Orientation) in the HE-AI model capture the often-overlooked individual's early engagement with AI, where curiosity, novelty, and low-barrier tools drive foundational literacy. The hybrid needs- and capability-based model prioritises the foundational psychological and educational (skills) dimensions as central to AI engagement.

PwC's and Deloitte's frameworks place scalable innovation at the top of AI maturity. For example, in Deloitte's framework, organisations that are in the pinnacle Transforming stage are "using AI pervasively and at scale," leading to new business models and market leadership (Deloitte, 2021). In PwC's framework, the top Transformational level encompasses the enterprise-wide deployment of AI across functions and services (PwC, 2020). The HE-AI Model goes a step further and incorporates scale and intentionality (Level 6 – Human-AI Co-Evolution) and contribution to foundational knowledge (Level 7 – Societal and Global Integration). At Level 7, organisations contribute to the future of AI through research leadership, global collaboration, and stewardship of AI for the public good. This aligns with "transcendence" in Maslow's hierarchy (Maslow, 1970). Butler et al.'s (2024) AI Capability Assessment Model (AI-CAM) and the AI Capabilities Matrix (AI-CM) focus on organisational readiness capability deployment. They provide practical tools for assessing an organisation's ability to understand and adopt AI. Their five-level framework (ad hoc adoption to enterprise-wide integration) is based on seven capability dimensions (business, data, technology, organisation, AI skills, risks, and ethics), emphasising the centrality of data governance, architecture and semantic technologies for sustainable AI deployment.

Trust and ethics-focused frameworks have also emerged in response to AI's influence in sensitive domains. Vakkuri et al. (2021) recognised the tension between ethics-focused approaches to AI development and more general AI maturity frameworks that integrate ethics as one dimension among others. They posit that AI (ethics) MM should be built on the software engineering MM tradition and embed ethical considerations into organisational processes. The Steinmetz et al. (2025) *Trust Calibration Maturity Model* is built on five trust dimensions to assess AI systems (performance, bias, transparency, safety, and usability). Their model characterises the trustworthiness of AI systems. Mäntymäki et al.'s (2022) *Hourglass Model of Organisational AI Governance* emphasises governance and visualises ethics as a filter between business goals and technical solutions. Zuiderwijk et al. (2024) explicitly include "ethics and responsibility" as a key evaluation dimension in their model. Reuel et al. (2025) introduced a two-dimensional Responsible AI (RAI) maturity model. Their model differentiates between system-level risk mitigations (e.g., discrimination, reliability, privacy) and organisational processes (governance, risk management, monitoring, and control).

De-Oliveira et al.'s (2024) analysis of digital MMs found that over the last 10 years, AI has been integrated into them mainly for organisational benchmarking and decision-making. Emergent themes, such as ethics, governance, and transparency, remain on the periphery of most of these MMs. In contrast, the proposed HE-AI model embeds inclusivity and accessibility at Levels 0 (Initial Exposure and Curiosity) and 1 (Awareness and Orientation), and ethics at Level 5 (Responsible Deployment and Operational Impact) as essential stages in AI engagement.

There has also been a surge in the development of agentic AI systems that are capable of sustained collaboration with humans. To measure maturity in the domain, Li et al. (2024) proposed a four-level capability MM that would track an LLM's journey from basic use to orchestrating multi-agent workflows that support human-AI co-reasoning. Using their model, they were able to show that AI agents deployed to support research yielded significant efficiency gains and improvements in research output quality compared with human baselines. They emphasised, however, that MMs should integrate technical capability scaffolding and governance mechanisms to ensure AI agents remain transparent, auditable, and aligned with human values while becoming better collaborators.

The proposed HE-AI model makes several contributions to the academic literature. It presents a hierarchical, motivational and capability-driven structure that links psychological needs (curiosity, esteem, mastery, contribution) to technical and infrastructural engagement with AI. Individual and organisational progression are combined within the model, unlike typical organisational MMs.

## 3. HE-AI Model Development Methodology

### 3.1 Theoretical Foundations

Maslow's hierarchy of needs (Maslow, 1943; 1970) progresses through seven stages (physiological survival, safety, belonging, esteem, self-actualisation, transcendence), each highlighting the motivational drivers of behavioural change (see Figure 1). Due to its structured, intuitive progression, the framework finds widespread use (Kenrick et al., 2010). Critics, however, claim that the model may not be readily applicable in non-Western contexts because the framework was based on Western assumptions (Neher, 1991). Others claim that there is limited empirical evidence to support Maslow's sequential progression, arguing that human motivations are often pursued non-linearly, simultaneously, or according to contextual priorities (Wahba & Bridwell, 1976). The relevance of the model, however, endures, driven by its ability to intuitively communicate complex human developmental processes and is adopted as the conceptual scaffold to frame AI engagement in the HE-AI model.

In the HE-AI model, engagement with AI moves through a hierarchy of motivations, competencies, and capabilities, with each level building on the previously established foundations. At lower levels, individuals and organisations satisfy basic needs, including exposure, literacy, and trust. They then move on to more complex aspirations, including mastering infrastructure, driving innovation, and responsible deployment. At the highest level, individuals and organisations have an impact on the development and governance of AI. The model aligns motivational drivers (curiosity, safety, mastery, contribution) with the technical, organisational, and ethical dimensions of AI adoption. It emphasises that AI maturity is not just technical competence, but a broader developmental progression incorporating values, trust, and responsibility.

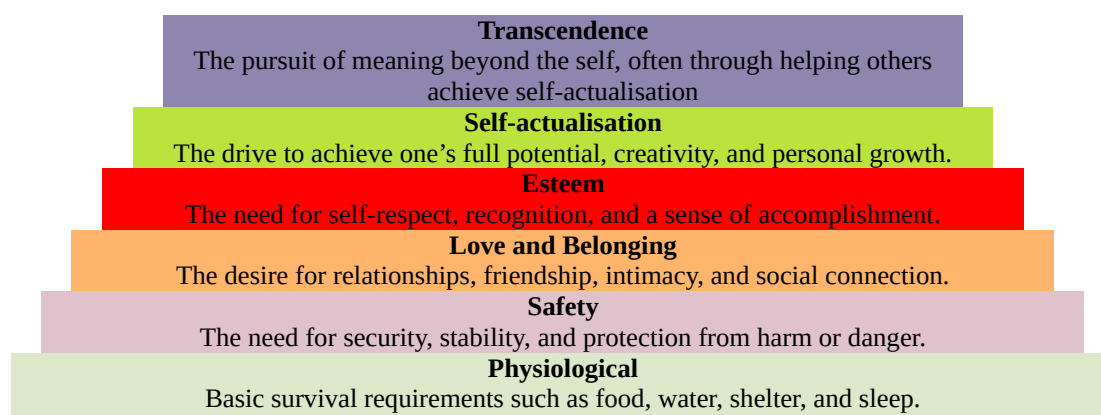

*Figure 1: Maslow's Hierarchy of Needs*

**3.2 HE-AI Model Development Stages**

The development of the HE-AI model followed a three-stage, iterative process aligned with established best practices in design science and MM construction (Becker et al., 2009; Gregor & Hevner, 2013; Mettler, 2011).

1. Conceptual framing. Inspired by Maslow's hierarchy of needs, AI engagement by individuals and in organisations was framed as an intuitive hierarchical construct that progresses from basic exposure to higher-order levels of responsibility, scaling, and ecosystem contribution. The framing is multi-dimensional, including technical capability, organisational processes, ethical governance, and societal impact.

2. Model structuring. The hierarchy was based on sequential levels, reaching eight levels (0–7) through a series of reflections and refinements to ensure internal coherence and consistency. Each level is defined by key characteristics, infrastructure requirements, the complexity of activities, the required skill levels, and the scope of the applications. This approach was based on MM design principles that stress the importance of providing clarity, progression, and internal coherence (Fraser et al., 2002; De Bruin & Rosemann, 2005). Early drafts focused on delineating clear transitions between levels based on existing models and the assessment of private and public sector organisations' AI journeys.

3. Validation through use cases. The applicability and robustness of the model were tested using four diverse validation cases: General Motors (industry), the government of Estonia (national government), the University of Texas System (higher education), and the African Union AI Strategy (continental policy). The organisations were chosen because of their differences, and their extensive user journeys are well documented and available in the public domain. Initiatives and activities from each case were mapped onto the HE-AI model to assess correspondence and identify ambiguities and gaps. Insights from these validations guided further model refinement.

The following section presents the HE-AI model.

**4. The HE-AI Model**

The HE-AI model is an eight-level individual and organisational journey of engagement with AI, characterised by increasing complexity, necessary skills, infrastructure requirements, and societal impact (see Figure 2). It starts at the level of basic exposure and trust-building, prerequisites for higher levels of innovation, responsibility, and contribution. Additionally, lower-level activities persist even when higher levels are achieved. For example, advanced researchers (at a higher level) may still experiment casually with generative tools (at a lower level), or institutions that have scaled (at a higher level) will still offer training workshops for their staff (at a lower level). The model is intended for wide use, for example, by individuals gaining AI literacy, public and private sector organisations embedding AI into their workflows, and policymakers designing national or continental strategies. It is positioned as a diagnostic tool ("where am I now?") and a prescriptive framework ("what do I need to do next?"). A brief outline of the eight levels of the model is presented in the sections that follow.

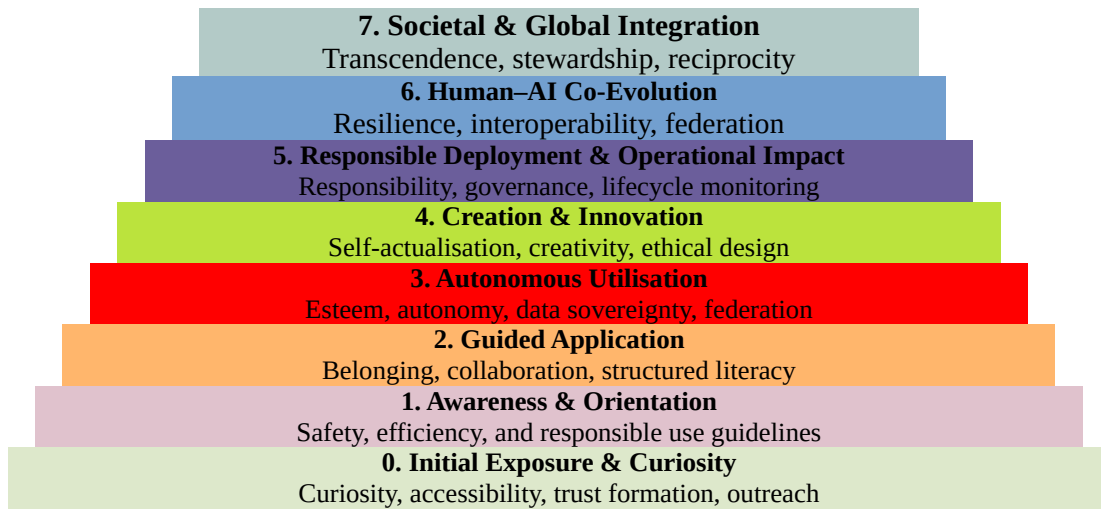

*Figure 2: The Hierarchy of Engagement with AI Model*

### 4.1 The HE-AI Model Levels

### 4.1.1 Level 0: Initial Exposure and Curiosity

Level 0 is the start of a user or organisation's journey of engagement with AI, driven by curiosity and exploration, using free or low-cost tools. Peer influence, institutional campaigns, or the media are some of the triggers for initial engagement at this level. For example, individuals may experiment with ChatGPT or DeepSeek for the first time, while governments or universities may run public awareness campaigns to introduce the public or students to AI. Level 0 is analogous to Maslow's "physiological needs" and provides the experiential base upon which higher levels are built. It forms users' first impressions of AI's usefulness, reliability, and fairness, which influences their likelihood of adoption. Key Level 0 motivations include curiosity and awareness expressed through trying AI for the first time; accessibility and inclusion through free, simple tools with low entry barriers; foundational literacy and trust formation as users begin to understand the capabilities, limits, and reliability of AI; policy outreach and support, through awareness campaigns and programmes; and non-linear persistence through recurring Level 0 activities to accommodate new cohorts engagement with AI.

### 4.1.2 Level 1: Awareness and Orientation

At Level 1, individuals and organisations shift from the broad curiosity of Level 0 to the intentional use of AI for specific, simple tasks. For example, the use of Grammarly for grammar correction, scheduling assistants, or chatbots for FAQs. Other examples include the use of paid premium levels of tools, such as ChatGPT, to access advanced features and functionalities; a factory implementing a simple AI-powered alert system on its production floor or an accountant using an internal AI-based expense categorisation assistant. Level is characterised by controlled, bounded applications that provide small productivity gains. It aligns with Maslow's "safety needs," and gives confidence in AI's usefulness. Formal and informal norms regarding the appropriate use of AI begin to be adopted. Key Level 1 motivations include task automation, where AI is applied to simple, repetitive tasks; assistance and support, with AI supporting drafting, translation, or scheduling; confidence achieved through consistent outputs; oversight through human review; and responsible use guidelines where individuals or organisations develop acceptable use norms. As confidence and oversight grow, engagement evolves into embedding AI into structured workflows and collaborative contexts at Level 2.

**Table 1: Level 1 Key Characteristics**

| Key Characteristics |
|---|
| • Introduction of department-level or pilot program-specific tools, e.g., simple chatbots, basic predictive analytics dashboards.<br>• Initial training focused on understanding and using these specific applications, low technical complexity ("low-barrier" tools).<br>• Basic understanding and orientation towards a limited set of defined AI functions. |

### 4.1.3 Level 2: Guided Application

Level 2 is a significant step beyond the initial awareness and explorations at Level 0 or the targeted application focus at Level 1. Example Level 2 applications include a university introducing plagiarism-detection software across the institution and providing training for lecturers on its use (Level 0 activity). A hospital chain may implement an AI-supported triage platform, requiring medical staff and the IT teams to work together. Level 2 activities use third-party infrastructure and platforms in addition to introducing formal training, organisational guidelines, and institutional oversight. These activities are analogous to Maslow's "belonging needs," as the use of AI transitions from the individual into the organisation. Key Level 2 motivations include structured workflow integrations, collaboration and shared practice (institutional teams adopt AI), training programmes to build capacity in AI, and ensuring responsible use and addressing reliability issues by introducing early governance with oversight mechanisms. Non-linear engagement continues through lower-level activities.

**Table 2: Level 2 Key Characteristics**

| Key Characteristics |
|---|
| • Embedding third-party AI tools/platforms into core operational processes.<br>• Development of internal guidelines for effective use of AI.<br>• Establishment of institutional oversight mechanisms to monitor AI use and manage associated risks or questions.<br>• Focus on practical implementation within defined structures.<br>• Early AI pilots reveal data management and governance gaps, leading to documentation of ownership, quality controls, and compliance requirements for future scaling (Butler et al., 2024).<br>• Identification of organisational workflows for embedding AI tools (e.g., customer service, content moderation, administrative tasks).<br>• Training programs tailored for user groups interacting with the new AI systems.<br>• Establishment of collaborative teams to manage the use of Level 2 AI assets. |

### 4.1.4 Level 3: Autonomous Utilisation

Level 3 moves beyond the initial integration into workflows of Level 2. Organisations transition from relying on third-party platforms or basic internal tools to actively building, controlling, and mastering their own dedicated AI infrastructure, including secure data pipelines. For example, governments building national digital public infrastructure (DPI) (e.g., Estonia's X-Road platform) to securely manage and share data across agencies, enabling sovereign AI deployment for critical functions; or large corporations procuring their own high-performance computing platforms that can run complex generative AI models or perform real-time data analytics. Level 3 activities develop autonomy, competence, and recognition analogous to Maslow's "esteem needs." Key Level 3 motivations include having and running one's own infrastructure, where secure and scalable AI environments are established; data governance and sovereignty as strategies for compliance with laws and regulations on data protection; and federation and interoperability of AI systems to support multi-institutional coordination of infrastructure. Non-linear engagement continues through lower-level activities.

**Table 3: Level 3 Key Characteristics**

| Key Characteristics |
|---|
| • Development or procurement of enterprise-grade platforms, e.g., internal cloud environments, high-performance computing clusters.<br>• Implementation of robust, secure data pipelines tailored to organisational needs.<br>• Formulation of roadmaps for AI adoption, linking business strategy, data governance, and skills planning (Butler et al., 2024).<br>• Mastery of secure data governance practices for sensitive organisational datasets.<br>• Owning critical infrastructure components. |

### 4.1.5 Level 4: Creation and Innovation

Organisations' focus shifts from building foundational platforms (Level 3) to developing tailored AI solutions. For example, governments may develop country-wide AI-enabled services running on national DPIs built at Level 3, or universities may develop specialised LLMs for specific researcher or educational needs. Creativity and problem-solving dominate these activities, making Level 4 analogous to Maslow's concept of "self-actualisation." Key motivations at this level include innovation and creativity, where novel AI tools are developed, tailored to local contexts, cross-disciplinary collaboration to advance research and practice, ethical design that embeds fairness and transparency into solutions, and contextual relevance, ensuring new models and solutions align with organisational missions and values. Non-linear engagement endures as lower-level activities continue.

**Table 4: Level 4 Key Characteristics**

| Key Characteristics |
|---|
| • Development of specialised solutions to address unique challenges or opportunities, e.g., sector-specific models, bespoke tools.<br>• Embedding AI across organisational functions beyond simple workflow integration.<br>• Cross-disciplinary collaboration to advance research and practice.<br>• Development of complex multi-agent workflows (Li et al., 2024).<br>• Ensuring robustness and reliability across diverse applications developed for different contexts or purposes.<br>• Integrating ethical considerations into the design process. |

### 4.1.6 Level 5: Responsible Deployment and Operational Impact

Level 5 is a critical transition point beyond the context-specific innovation focus of Level 4. Organisations shift from developing novel AI solutions to implementing them into mission-critical contexts (e.g. autonomous vehicles, healthcare, finance, government services) that are underpinned by external governance and oversight. For example, governments deploying AI-enabled public services at scale with embedded accountability systems and ongoing monitoring. Level 5 serves as the "ethics gate" where only responsibly governed systems scale. It reflects Maslow's higher-order "being-values" that emphasise truth, fairness, and integrity. Key motivations at this stage include governance and ethics where accountability is embedded into AI use, transparency as systems are put in place to ensure AI decisions can be audited and understood, adding a human-in-the-loop oversight so that critical review remains with people, lifecycle monitoring to detect drift, bias, and unintended effects within the AI systems, and building public trust through regulatory alignment and compliance. Non-linear engagement continues through lower-level activities.

An emerging responsibility at Level 5 involves data provenance and licensing. Provenance refers to the origin, history, and conditions of use of training data, i.e., where it came from, how it was obtained, and whether its use is legally and ethically permissible. Organisations must demonstrate that the training data used in their AI systems was sourced transparently and ethically. Empirical studies show that some advanced LLMs were trained on non-public, copyrighted data without consent, raising legal and sustainability concerns (Rosenblat et al., 2025). Incorporating provenance checks, licensing mechanisms, and disclosure of data sources into deployment practices strengthens public trust and provides a measurable gate before moving to higher maturity levels.

Level 5 also signifies the onset of post-deployment observability. Recognising that pre-deployment evaluations, such as alignment and testing, are important, they do not adequately capture the risks that emerge once systems are deployed in real-world contexts. Organisations prioritise pre-deployment efforts, leaving gaps in understanding potential harmful impacts during deployment, such as misinformation, addictive design, copyright violations, and hallucinations (Strauss et al., 2025). Organisations at Level should implement structured monitoring of deployed systems, strengthening accountability and ensuring operational impacts are transparent and responsive to societal risks.

**Table 5: Level 5 Key Characteristics**

| Key Characteristics |
|---|
| • Deployment of solutions developed at Level 4 into mission-critical operational contexts requiring high reliability and safety standards.<br>• Establishment of comprehensive governance frameworks covering ethical AI use (embedding ethical guardrails), risk mitigation for sensitive areas, bias auditing, transparency, and accountability mechanisms.<br>• Implementation of lifecycle monitoring systems to track performance, detect drift or unintended effects in scaled-up operations.<br>• Building public trust through demonstrable compliance with regulations and robust oversight structures.<br>• Transparent documentation of AI-assisted processes and maintaining human accountability for outputs, ensuring that AI enhances, not replaces, human responsibility (Li et al., 2024).<br>• Routine measurement of return on investment, cost–benefit, and risk, supported by continuous data governance processes (Butler et al., 2024).<br>• Conducting ongoing audits for bias, fairness, and performance in operational settings. |

### 4.1.7 Level 6: Human–AI Co-Evolution

Level 6 embodies the system-wide scaling of AI systems, accompanied by the necessary ethical guardrails and governance structures that were established in Level 5. For example, governments may implement country-wide AI-enabled public services with embedded accountability systems, ongoing monitoring for performance drift/bias, and transparent oversight mechanisms; or scaling federated data platforms for use in disease surveillance or personalised medicine across multiple hospitals with data sovereignty rules, privacy safeguards, and collaborative lifecycle monitoring. Level 6 reflects Maslow-inspired "organisational self-actualisation at scale." Key motivations at this stage include durable scaling where AI is embedded across entire organisations or ecosystems, reliability and resilience reflecting the necessity of the AI systems to be adaptable and perform consistently, federation and interoperability supporting the coordination of standards across organisations or regions, and workforce pipelines to develop the skills necessary to sustain long-term adoption. Non-linear engagement continues through lower-level activities.

**Table 6: Level 6 Key Characteristics**

| Key Characteristics |
|---|
| • Achieving reliable and safe AI deployment across complex operational environments.<br>• Establishing robust frameworks for ongoing transparency, accountability, bias auditing, and performance monitoring at scale.<br>• Developing capabilities for seamless interoperability with external partners or systems supporting coordinated deployment between organisations.<br>• Developing specialised workforce competencies to manage sophisticated AI deployments.<br>• Standardising post-deployment evaluation hooks and telemetry for agentic systems, enabling scaled AI-supported workflows to be independently audited and compared across organisations (Li et al., 2024). |

At scale, federated data governance frameworks become essential. Just as interoperability standards enable technical resilience, federated mechanisms for data licensing, consent management, and provenance auditing enable ethical resilience across organisations and regions. Level 6 aligns with regulatory developments such as the European Union's AI Act, which requires general-use AI developers to provide "sufficiently detailed summaries" of training data (European Parliament, 2024). Embedding provenance into federated governance shall ensure that scaling reinforces trust and legality.

Also, post-deployment monitoring evolves into shared standards and collective practices. As organisations federate infrastructure and governance, they also federate observability frameworks that enable cross-institutional monitoring of AI behaviour, including adoption of standard telemetry formats, structured incident reporting, and interoperable logging systems that support independent research and audit (Strauss et al., 2025). These practices mirror the evolution of security and quality standards in other regulated industries and ensures that AI ecosystems mature with technical resilience and ethical accountability.

### 4.1.8 Level 7: Societal and Global Integration

Level 7 represents the highest level of AI engagement maturity. Organisations contribute to shaping global standards and achieving widespread societal adoption through reliable, interoperable, and ethically guided AI systems. Organisations may release open-source AI platforms, contribute to global standards, or build cross-border interoperable systems. For example, the Government of Estonia open-sourcing its AI Assistant and AI companies making available open-source versions of their LLMs. Level 7 corresponds to Maslow's concept of "self-transcendence" and focuses on stewardship and reciprocity. Key motivations include ecosystem collaboration with organisations and governments to build mutually-beneficial partnerships; making contributions to global standards, international norms and governance; open contributions such as releasing datasets, platforms or models; having societal-level impact through AI advances in diverse sectors (e.g. education, health); supporting interoperability to ensure AI systems integrate across jurisdictions; and reciprocity and stewardship.

Level 7 AI actors move beyond compliance to actively shaping global data commons, for example, by supporting international standards for provenance disclosure, contributing to licensing marketplaces that fairly compensate content creators, and stewarding transparent, open datasets that serve as public goods for AI training (Rosenblat et al., 2025). The activities ensure that organisations deploying AI systems contribute to the sustainability of the very ecosystems of knowledge and creativity upon which they depend. Responsibility also extends to the global governance of deployment-stage risks, including contributing to international monitoring frameworks, reinforcing reciprocity and trust, and ensuring that AI systems evolve in a manner that benefits society and is governed sustainably (Strauss et al., 2025).

The descriptions of Levels 0–7 of the HE-AI model illustrate the structured progression pathway from initial AI exposure to ecosystem-level contribution provided by the HE-AI model. Table 8 highlights the differentiating characteristics between the model levels. The following section presents the validation cases used to test and refine the HE-AI model.

**Table 7: Level 7 Key Characteristics**

| Key Characteristics |
|---|
| • Contributing to the development of AI models through open datasets, models and platforms.<br>• Contributing to the development of international frameworks for responsible AI deployment and operation.<br>• Designing governance frameworks capable of coordinating ethical oversight across multiple jurisdictions. |

## 4.2 Validation Use Cases

A structured six-step case analysis methodology was used to validate and refine the HE-AI model, testing the model's applicability across four diverse domains and identifying areas that needed strengthening:

(a) Identification of diverse illustrative organisations;

(b) Development of detailed and properly referenced case studies from publicly available information using ChatGPT;

(c) Systematic review of each case to extract key AI-related activities/initiatives;

(d) Mapping activities/initiatives onto the HE-AI model;

(e) Identification of activities/initiatives that do not align with existing levels (these were treated as signals of potential gaps or areas requiring further model refinement); and

(f) Identification of ambiguous or dual-level placement of activities/initiatives (evidence of weak distinctions between levels and opportunities to strengthen definitions).

For the policy case, the African Union AI Strategy (African Union, 2024) served as the primary source, followed by steps (c) to (f). The approach follows MM design and validation best practices (Proença & Borbinha, 2018; Venable et al., 2016).

The presentation of the four validation and refinement use cases follows.

### 4.2.1 General Motors

General Motors (GM) demonstrates the progression through multiple levels of the HE-AI model as a large multinational company. As presented in Table 9, GM's journey spans from early exploratory design and manufacturing pilots (Levels 2–3) (Fearn, 2025), to significant investments in autonomous vehicle innovations through the acquisition of Cruise Automation, a company that was developing autonomous vehicle technology, and investment in Lyft, a company working on autonomous taxis (Level 4) (White, 2016). AI engagement also included the development of technologies for safety-critical systems such as Super Cruise (Level 5) (Voelcker, 2023). GM's subsequent scaling of AI-based operations across its global manufacturing facilities (Clausen, 2025) demonstrates Level 6, with movement to Level 7 as the company embarked on initiatives in electronic vehicle battery research and cross-industry collaborations (General Motors, 2025). The GM case led to several significant refinements of the HE-AI model. The case demonstrated the persistence of lower-level activities, assets and capabilities even as higher levels are reached, suggesting model non-linearity. Additionally, it highlighted the need to incorporate an ethics and governance "gate" between Levels 5 and 6, as demonstrated by the regulatory setbacks the company experienced in its Cruise's robotaxi programme (Shepardson & Eckert, 2024). In addition, GM's experience with AI revealed that both economic viability (for commercial organisations) and public trust in the AI-enabled products are critical to progression along the HE-AI model alongside technical capacity.

### 4.2.2 Government of Estonia

The Estonian government's early projects, such as AI-enabled traffic policing, anomaly detection on X-Road, and debt-claim automation, can be situated at Levels 2–4 (see Table 10), where AI is embedded into workflows and bespoke domain-specific applications were created (European Commission AI Watch, 2024; e-Estonia, 2025). The rapid deployment of the Suve chatbot during the COVID-19 pandemic (Levels 2–3) provided scalable public information services in a crisis (Schwartz, 2020). By 2021, the launch of Bürokratt, a nationwide AI assistant with embedded human oversight, marked Estonia's progression into Level 5 (OSOR, 2022). Notably, the rejection of a proposed "robot judge" highlighted Level 5's ethics gate, ensuring that final discretion in justice remained with humans (Estonian Ministry of Justice, 2022). Estonia has since progressed to Level 6 through system-wide initiatives, including a cross-government AI-powered data management platform (Vincent, 2025) and the AI Leap 2025 programme, which introduced AI tools and training to all upper-secondary schools (Weale, 2025). Emerging Level 7 behaviours are visible in Estonia's open-sourcing of the Bürokratt platform for international adoption and its contributions to global AI governance discussions (OSOR, 2022).

The Estonia case led to the following refinements of the HE-AI model. First, it highlighted the critical role of pre-existing DPI, such as eID and X-Road, as a foundational and enabling factor for progression to higher levels. It also underscored that regulatory co-evolution and public trust are continuous, cross-level requirements rather than confined to a single stage. Finally, Estonia's experience highlighted the importance of partnership maturity and low-resource language adaptation as key factors influencing AI adoption in smaller states.

### 4.2.3 University of Texas System

The University of Texas (UT) System illustrates how an extensive federated higher education network progresses across multiple HE-AI levels. As shown in Table 11, Levels 0–2 were supported through initiatives such as UT Austin's "Year of AI," which promoted curiosity and cross-disciplinary engagement (University of Texas at Austin, 2024a), and UT Tyler's syllabus policies embedding responsible use into early classroom practice (University of Texas at Tyler, 2024). Level 3 initiatives included the establishment of the Texas Advanced Computing Centre (TACC) and the Frontera supercomputer, which provided secure, large-scale infrastructure for AI research (National Science Foundation, 2019). At Level 4, innovation was demonstrated through UT Austin's leadership of the NSF Institute for Foundations of Machine Learning (IFML) and UT Dallas's launch of its Institute for AI (University of Texas at Dallas, 2025).

Responsible deployment at Level 5 was identified in programmes such as the joint M.D./M.S. in AI at UT San Antonio and UT Health San Antonio, which integrate ethics and human oversight into clinical training (University of Texas at San Antonio, 2023). UTSA's creation of a College of AI, Cyber and Computing, consolidating over 5,000 students into a durable academic structure (University of Texas at San Antonio, 2024a), alongside TACC's sustained provision of advanced infrastructure, demonstrated scaling at Level 6. Finally, Level 7 was evidenced through initiatives such as the Good Systems program and UT's leadership roles in international AI ethics consortia (University of Texas at Austin, 2024b). The UT case prompted key refinements to the HE-AI model: the importance of federated governance across multiple institutions, the integration of workforce development pipelines, and the role of stage-gates between responsible deployment (Level 5) and durable scaling (Level 6).

**4.2.4 African Union AI Strategy**

The African Union's (AU) Continental AI Strategy (African Union, 2024) demonstrates how a continental policy framework aligns with the HE-AI model. As shown in Table 12, Levels 0–2 are reflected in commitments to AI literacy and media and information literacy for youth and educators. Level 3 is evident in the establishment of regional centres of excellence and observatories to strengthen shared infrastructure and data governance. Level 4 ambitions are reflected in support for research, innovation, and startup ecosystems in priority sectors, including agriculture, health, education, and climate resilience. Central to the AU's strategy is the embedding of rights-based governance frameworks and ethical guardrails to ensure responsible deployment and prevent bias, discrimination, and misuse (Level 5). At Level 6, the strategy emphasises scaling through continental investments in broadband, electricity, data centres, and cloud infrastructure, supported by workforce training pipelines. Finally, the AU's objective to influence global AI governance, promote cross-border interoperability, and protect data sovereignty, while positioning Africa as a global stakeholder, falls within Level 7. The AU case led to key refinements of the HE-AI model, highlighting the importance of multi-level governance topologies (continental, regional, and national), the role of resource mobilisation and funding readiness as enablers of scaling, and the need to account for information integrity and indigenous knowledge protection in higher-level transitions.

The four validation cases identified areas that informed model refinement and demonstrated the adaptability of the HE-AI model across multiple contexts. They confirm core elements of the HE-AI model, including non-linearity, persistence of lower-level activities, and the centrality of responsible deployment. The selected cases, however, were from large, resource-rich institutions and national or continental strategies. To further strengthen the robustness of the HE-AI framework, future work shall investigate use cases from small- and medium-sized enterprises, non-governmental organisations, and organisations that operate with limited infrastructure and funding. SMEs often face barriers to AI adoption, including a lack of expertise, limited data resources, and constrained budgets (Dubey et al., 2022; Pradhan et al., 2023). Similarly, NGOs and community-based organisations frequently engage with AI under conditions of informality and improvisation, prioritising social value creation over efficiency or competitiveness (Starke et al., 2022). Additionally, Global South organisations often face infrastructural deficits, skills gaps, and uneven policy environments, leading to varying AI adoption trajectories (Adam et al., 2020; Tiseo et al., 2023). Incorporating insights from these environments shall extend the model's diagnostic utility and strengthen its objective to be inclusive and adaptable to a broader diversity of AI engagement journeys.

It is important to properly situate the HE-AI model by comparing it against existing AI maturity and capability frameworks, many of which have emerged from corporate, academic, and policy environments. The following section, therefore, positions the HE-AI model within the broader landscape of maturity models, clarifying how it complements these global approaches.

### 4.3. Positioning the HE-AI Model Within Existing AI Maturity Frameworks

The HE-AI model builds on, but also departs from, established AI maturity frameworks by reframing maturity as a progression driven by human motivations, trust, ethics, and societal contribution, rather than only technical or organisational readiness. Traditional models such as PwC's *AI Capability Maturity Model* (PwC, 2020), Deloitte's *AI Maturity Framework* (Deloitte, 2021), Gartner's *AI Maturity Curve*, and the *MIT CISR Enterprise AI Maturity Model* (Weill et al., 2024) mainly focus on how organisations develop infrastructure, governance, and enterprise capabilities. These models provide valuable roadmaps for organisations to scale their engagement with AI. They typically take AI awareness as a given, the adequate resources being available, and evaluate success mainly in terms of operational efficiency, competitive advantage, or enterprise-wide integration.

The HE-AI model complements these approaches by adding underrepresented dimensions. It recognises initial exposure and trust-building (Level 0), lowering the entry barrier for individuals, organisations and the general public where AI literacy cannot be assumed. The model introduces an ethics "gate" at Level 5, emphasising lifecycle monitoring, transparency, and governance as prerequisites for scaling, rather than as parallel or optional activities. Also, it stresses federation, resilience, and interoperability at Level 6, acknowledging the collaborative and multi-level governance realities of education systems, governments, and regional networks. Finally, at Level 7, the model reframes maturity as societal and global integration, highlighting stewardship, reciprocity, and cross-border interoperability. With reference to Table 13, models such as PwC, Deloitte, Gartner and CISR do not extend to motivational entry points, ethics as a gate, or global stewardship. The HE-AI model thus fills a crucial gap by offering a framework that includes individual and policy contexts, incorporates ethical progression, and captures the societal-level impact.

Reuel et al.'s (2025) Responsible AI maturity model distinguishes between system-level measures, such as fairness, reliability, and privacy controls, and organisational processes, including governance, monitoring, and risk management. Their model presents the steps organisations could take to close the gap between stated commitments and operational practice. The HE-AI model extends this work by embedding responsible AI not as parallel dimensions but as stage-gated requirements for organisational progress through successive levels of maturity, directly addressing the planning–execution gap (Vakkuri et al., 2021; Reuel et al., 2025). This model also addresses the concern that high-level principles lack enforceability (Mittelstadt, 2019) by providing the structure to ensure that ethical governance is not marginalised within broader AI development.

### 5. Discussion and Conclusions

The Hierarchy of Engagement with AI model provides a structured framework for understanding how individuals and organisations progressively deepen their engagement with AI. Based on the conceptual scaffolding of Maslow's hierarchy of needs, the model places AI engagement along a spectrum starting with basic exposure and culminating in ecosystem collaboration and societal impact. This conceptual foundation yields a model that is both intuitive and adaptable. The sequential levels of the HE-AI model capture the multiple dimensions of AI engagement. Each level represents increasing technical sophistication coupled with expanding social, ethical, and organisational commitments. The analogy with Maslow's hierarchy establishes a coherent progression from basic survival needs (exposure) to safety and belonging (early adoption and collaboration), then to esteem (infrastructure mastery), self-actualisation (innovation and responsible practice), and ultimately self-transcendence (ecosystem stewardship). This progression illustrates that AI maturity, both technical and human, is rooted in values, trust, and social responsibility.

The validation use cases were central to testing and refining the model. Their diverse contexts demonstrate that the HE-AI levels are observable in practice across industry, government, educational and policy ecosystems. Together, these cases demonstrate the flexibility of the HE-AI model. The HE-AI model provides a comprehensive and actionable framework for understanding and guiding AI maturity. It demonstrates that AI engagement is multi-dimensional and that engagement progression requires both structural foundations and value-driven governance. The model enables individuals, organisations, and governments to assess their current position on their AI journeys and to chart sustainable pathways forward.

**7.0 Limitations and Future Work**

While the proposed HE-AI model offers a structured framework for understanding how individuals and institutions progress in their interaction with AI systems, its contribution ultimately depends on demonstrating empirical robustness and practical relevance beyond the four use cases. As the model integrates technical, organisational, and ethical dimensions in ways not fully captured by existing frameworks, it is critical to validate its accuracy, usability, and adaptability across diverse contexts. To ensure the robustness, credibility, and practical applicability of the proposed HE-AI model, future work shall employ a structured multi-phase validation strategy. The approach will combine theoretical assessment, institutional piloting, cross-framework benchmarking, and empirical stakeholder feedback, ensuring that the model accurately represents the progressive stages of AI engagement while remaining adaptable to diverse organisational and cultural contexts (CMMI Institute, 2020; Deloitte, 2021; PwC, 2020).

Expert reviews and theoretical validation shall be performed by engaging specialists in AI ethics, digital transformation, organisational innovation, and policy. These experts will evaluate the conceptual clarity of each level and assess alignment with existing maturity models. Also, a large and diverse set of additional cases will be developed to validate further and refine the model. Empirical stakeholder engagement will be conducted through surveys and structured interviews with diverse stakeholders to assess the model's comprehensiveness, usability, and the extent to which it accurately captures real-world AI engagement journeys. The collected information will directly inform iterative refinements to ensure that the model remains relevant as AI capabilities, governance frameworks, and societal priorities evolve.

The development of indicators that make the AI engagement journey progression across levels observable and measurable is an essential step in operationalising the HE-AI model. The availability of indicators enables the model to function as both a diagnostic tool for assessing current AI engagement and a prescriptive framework for guiding advancement. Indicators also create a shared vocabulary for comparison, enabling benchmarking across institutions, sectors, or national contexts. Future work will develop generic and sector-specific indicators. Generic indicators capture broad patterns of progression across levels. For example, measures of AI literacy and training penetration (Levels 0–2), the existence of governance and data protection frameworks (Levels 3–5), or the degree of system-wide interoperability and scaling (Levels 6–7). These indicators provide a useful high-level view, but may be too abstract to capture the nuanced realities of different domains. For example, "AI literacy" will manifest differently in a university compared to a government ministry or a manufacturing company.

Illustrative sector-specific indicators were developed for the four use cases (see Table 14). They demonstrate how HE-AI levels can be contextualised to reflect distinct priorities, infrastructures, and governance challenges across different domains. The strength of sector-specific indicators is their ability to highlight differences in pathways and bottlenecks. For example, for the UT systems case (higher education), progression hinged on capacity-building, workforce pipelines, and federated governance across multiple campuses, whereas in the Estonia case (public sector), success was contingent on digital public infrastructure and citizen trust. For GM (private sector), indicators focused more heavily on portfolio governance, lifecycle monitoring, and customer trust, while for the African Union (continental policy), resource mobilisation and cross-border interoperability were defining factors. These were adequately captured by the sector-specific indications. By tailoring indicators to sectoral realities, therefore, the HE-AI model becomes a more useful tool for diagnosis and strategic planning. Future work will extend the development of indicators and create corresponding scoring rubrics.

**Table 8 – HE-AI Model: Key Characteristics at Each Level**

| HE-AI Level | Motivational Drivers | Infrastructure Requirement | Complexity of Engagement | Primary Scope of Application | Maslow Analogy |
|---|---|---|---|---|---|
| **0. Initial Exposure & Curiosity** | Curiosity, accessibility, trust formation, outreach | None (free/public tools, demos) | Low – exploratory, informal use | Individuals, outreach programmes | Physiological Needs - Provides the experiential "survival base" for engagement; without exposure, higher engagement is impossible. |
| **1. Awareness & Orientation** | Safety, efficiency, responsible use guidelines | Minimal (consumer/enterprise apps) | Low–moderate – bounded tasks | Individuals, small teams | Safety Needs - Builds stability and trust in AI's reliability while embedding early safeguards. |
| **2. Guided Application** | Belonging, collaboration, structured literacy | Institutional licences, training platforms | Moderate – workflow integration | Teams, organisations (early) | Belonging Needs - Reflects community, collaboration, and reinforcement of AI adoption within groups and institutions. |
| **3. Autonomous Utilisation** | Esteem, autonomy, data sovereignty, federation | Organisational infrastructure, secure data pipelines | Moderate–high – independent deployment | Organisations, governments | Esteem Needs - Creates confidence, recognition, and independence through mastery of infrastructure and data. |
| **4. Creation & Innovation** | Self-actualisation, creativity, ethical design | R&D capacity, innovation labs, domain expertise | High – bespoke model development | Organisations, research groups | Self-Actualisation - Represents creative fulfilment and innovation, where organisations express their potential through novel AI solutions. |
| **5. Responsible Deployment & Operational Impact** | Responsibility, governance, lifecycle monitoring | Regulated infrastructure, oversight systems | High – mission-critical applications | Enterprises, governments | Self-Actualisation (Checkpoint) - Ensures that creativity is translated into ethical, transparent, and trustworthy practice—the "being-values" of self-actualisation in action. |
| **6. Human–AI Co-Evolution** | Resilience, interoperability, federation | Advanced digital ecosystems, workforce pipelines | Very high – cross-organisational scaling | Systems, federated institutions | Organisational Self-Actualisation at Scale - Extends self-actualisation into durable, reliable, system-wide maturity where AI is embedded as a core capability. |
| **7. Societal & Global Integration** | Transcendence, stewardship, reciprocity | Global platforms, cross-border standards | Very high – international collaboration | Ecosystems, societies, global governance | Self-Transcendence - Moves beyond self-actualisation to stewardship and reciprocity, creating value for broader ecosystems and society. |

**Table 9: Mapping General Motors AI Journey to the HE-AI Model**

| Year(s) | GM Milestone / Activity | HE-AI Level | Rationale / Notes |
|---|---|---|---|
| 2014–2016 | Acquisition of Cruise Automation; $500M Lyft investment; launch of Maven | Level 4 | GM shifted from adopter to creator, developing proprietary AV solutions and mobility services (White, 2016). |
| 2015–2016 | Early machine learning pilots in design and manufacturing | Level 2–3 | AI applied to repeatable tasks and secure integrations in engineering and operations (Fearn, 2025). |
| 2017 | Launch of *Super Cruise* (hands-free highway driving) | Level 5 | Safety-critical deployment with driver monitoring, geofencing, and governance, embodying ethics-by-design (Voelcker, 2023). |
| 2017–2018 | Generative design (Autodesk partnership) | Level 4 → Level 5 | Creative AI (L4) transitioned into governed production use (L5) once designs were validated (DeNittis, 2023). |
| 2021 | Microsoft Azure partnership for Cruise and enterprise AI | Level 3 → Level 5 | Enterprise-scale infrastructure control (L3) enabling operational impact at scale (L5) (Lienert & Shivdas, 2021). |
| 2020–2022 | Digital twins and vision-based quality control in factories | Level 5 → Level 6 | Operational deployment preparing for scale, reliability, and observability (Clausen, 2025). |
| 2022–2023 | Cruise driverless robotaxi pilots (SF, Austin, Phoenix) | Level 6 | Multi-region deployments requiring resilience and CI/CD at scale; first true L6 external attempt (Shepardson & Eckert, 2024). |
| 2023 | Pedestrian incident in San Francisco; disclosure failures | L5→L6 (Breakdown) | Trust and governance lapses underscored fragility of Level 5→6 transitions (Shepardson & Eckert, 2024). |
| 2024 | Strategic retreat from robotaxi; reintegration into ADAS (Ultra Cruise) | Regression to Level 5 (with Retention) | Illustrates non-linear progression: scaling back L6 ambitions but reusing assets for safer L5 domains (Shepardson & Eckert, 2024). |
| 2025 | Scaled AI in manufacturing: predictive maintenance, digital twins, ergonomics analysis | Level 6 | Demonstrates mature global deployment with observability and resilience (Clausen, 2025). |
| 2023–2025 | Investment in Mitra Chem (AI-accelerated EV battery discovery) | Early Level 7 | Cross-industry AI partnerships aimed at open innovation, though still emerging (General Motors, 2025). |

**Table 10: Mapping the Government of Estonia AI Journey to the HE-AI Model**

| Year(s) | Milestone / Activity | HE-AI Level | Rationale / Notes |
|---|---|---|---|
| 2017 | Proposal of "kratt law" to give AI systems legal personality | Level 0–1 | Sparked public/legal debate on AI ethics and liability; no implementation, but set the stage for national AI policy (Grigoryan, 2019). |
| 2018 | AI Task Force established; Digital Summit on AI governance | Level 2 | Structured engagement with experts; early pilots scoped; signaled international leadership (e-Estonia, 2018). |
| 2019 | Publication of *Kratt Report*; launch of National AI Strategy (2019–2021) with €10M funding | Level 3 ) | Formalised AI strategy with funding, datasets, and pilots across ministries; move from pilots to sustained programmes (European Commission AI Watch, 2024). |
| 2019–2020 | AI-enabled policing, X-Road anomaly detection, National Library auto-tagging, debt-claim automation | Level 3–4 | Domain-specific AI solutions commissioned by ministries; bespoke models for justice, policing, and cultural heritage (e-Estonia, 2025). |
| 2020 | COVID-19 "Suve" chatbot launched nationally in 10 days | Level 2–3 | Guided but autonomous AI in crisis response; increased trust in AI for citizen-facing services (Schwartz, 2020). |
| 2021 | Dozens of AI solutions in government (80+ projects in development); Chief Data Officer drives open data expansion | Level 3–4 | Broad uptake across agencies; systemic data governance improvements (GovInsider, 2019). |
| 2022 | Launch of **Bürokratt**, national AI virtual assistant platform; clarification that Estonia will not develop an "AI judge" | Level 5 | Ethics-by-design and human oversight prioritized; open-source platform for all citizens; ethical checkpoint before scale (OSOR, 2022). |
| 2022–2023 | Expansion of Bürokratt across agencies; public surveys show growing trust; 80+ AI projects in 40 institutions | Level 5 | Responsible deployment validated by user uptake and transparency practices (e-Estonia, 2023). |
| 2024 | 100% of government services available online; preparation for generative AI integration | Level 5 → 6 | National platform readiness for AI integration; proactive services scaled across agencies (e-Estonia, 2024). |
| 2025 | Launch of **AI Leap 2025** (AI tools & training in all high schools); deployment of cross-government AI data management tool | Level 6 | System-wide scaling of AI across education and public service data; demonstrates national-level AI infrastructure (Weale, 2025; Vincent, 2025). |
| 2025 | Bürokratt open-sourced globally; interoperability with Finland's AuroraAI explored | Early Level 7 | Contribution to international AI ecosystems; Estonia as a model for small-state AI governance (OSOR, 2022). |

**Table 11: Mapping the University of Texas System AI Journey to the HE-AI Model**

| Year(s) | UT Milestone / Activity | HE-AI Level | Rationale / Notes |
|---|---|---|---|
| 2019 | Launch of *Frontera,* fastest academic supercomputer in the U.S. (NSF-supported) | Level 3 | Demonstrates control of compute and data infrastructure, essential for AI scaling (National Science Foundation, 2019; Texas Advanced Computing Center, n.d.). |
| 2018–2020 | Establishment of Good Systems (grand challenge on ethical AI) and NSF IFML (AI Institute for Foundations of Machine Learning) | Level 4 | Development of new methods and theoretical advances in AI, alongside ethical frameworks (University of Texas at Austin, 2024b; Institute for Foundations of Machine Learning, n.d.). |
| 2023 | UTSA & UT Health San Antonio launch dual M.D./M.S. in Artificial Intelligence | Level 5 | Integrates AI into clinical training with explicit human oversight, embodying ethics-by-design (University of Texas at San Antonio, 2023). |
| 2024 | UT Austin declares "Year of AI" with seminars, showcases, and cross-campus events | Level 0–2 | Broad campus-wide awareness and curiosity, embedding early responsible practice (University of Texas at Austin, 2024a). |
| 2024 | UT Tyler introduces mandatory AI-use language in syllabi | Level 1 | Embeds ethical guidelines and transparency into everyday academic practice (University of Texas at Tyler, 2024). |
| 2024 | UT Dallas hosts *Week of AI*; UT System organizes "AI Across the UT System" webinars | Level 2 | Provides guided, repeatable applications of generative AI in pedagogy (University of Texas at Dallas, 2025; University of Texas at San Antonio, 2024b). |
| 2024 | UTSA announces new College of AI, Cyber and Computing (launching Fall 2025) | Level 6 | Academic restructuring to serve >5,000 students, signaling durable scaling of AI education (University of Texas at San Antonio, 2024a). |
| 2025 | UT faculty publish system-wide guidelines for responsible AI in teaching and learning; UT participates in cross-institutional AI ethics consortia | Level 5 → 7 Transition | Consolidates governance (L5) while contributing to societal/global AI ethics discourse (University of Texas at Austin, 2024b; University of Texas at Dallas, 2025). |

**Table 12: Mapping the African Union AI Strategy to the HE-AI Model**

| Year(s) | AU Milestone / Activity | HE-AI Level | Rationale / Notes |
|---|---|---|---|
| 2018–2020 (precursors) | Pan-African discussions on digital transformation; AU Data Policy Framework; Malabo Convention on Cybersecurity & Data Protection | Level 0 | Early awareness of digital/AI opportunities and initial regulatory baselines; essential groundwork for later AI adoption. |
| 2021–2023 | AU consultative processes on AI; collaboration with UNESCO on AI readiness assessments | Level 1–2 | Structured exploration, stakeholder consultations, and continental-level scoping studies. |
| July 2024 | Formal adoption of the *Continental AI Strategy* (Accra) with six focus areas and 15 recommendations | Level 3 | Strategy establishes regional centres of excellence, observatories, and AU advisory structures—cross-country collaboration central. |
| 2024–2025 | Emphasis on AI literacy, skills development, and Media & Information Literacy (MIL) in education; plans for curricula and youth training | Level 2–3 | Embeds guided exposure and early adoption across population segments, preparing for broader integration. |
| 2024–2025 | Startup ecosystem support; promotion of sectoral R&D in agriculture, education, health, climate, and peace & security | Level 4 | Encourages new, context-specific AI applications and entrepreneurship. |
| 2024–2027 | Governance and regulation frameworks: ethical, rights-based, risk-mitigating (bias, discrimination, disinformation, surveillance, privacy) | Level 5 | Ethics and regulation sit at the centre of the strategy, aligned with HE-AI's Level 5 "gate" before scaling. |
| 2024–2030 | Capacity building in infrastructure: power, broadband, data centres, compute, and high-quality datasets; talent pipelines | Level 6 | Explicit focus on scaling reliable infrastructure and data capacity across the continent. |
| 2025–2030 | Implementation roadmap: phased plan with 2027 mid-term review; AU-level monitoring portal and readiness index | Level 6 | Mirrors HE-AI's emphasis on continuous monitoring and scaling resilience. |
| 2025 onward | Continental/global governance engagement; regional/international cooperation; shaping global AI standards | Level 7 | AU positions itself as a collective voice in global AI governance and standard-setting. |

**Table 13: Sector-Specific Illustrative Indicators of Engagement at Each HE-AI Level**

| Level | GM (Industry) | Estonia (National Government) | AU AI Strategy (Continental Policy) | UT System (Higher Education) |
|---|---|---|---|---|
| **0 – Initial Exposure & Curiosity** | % of employees exposed to AI pilots or demos; No. of internal awareness workshops. | No. of citizens engaging with national AI awareness tools (e.g., Suve chatbot usage stats). | No. of AI literacy and media & information literacy (MIL) campaigns launched across AU member states. | No. of students/faculty attending AI introduction events (e.g., "Year of AI" participation). |
| **1 – Foundational Task Automation** | % of supply chain processes automated with AI; error-reduction rates from automation. | % of administrative tasks automated using AI; average processing time reduction. | % of AU secretariat back-office processes supported by AI tools. | % of courses with AI-use policies; proportion of students using AI for basic academic support tasks. |
| **2 – Complex Application** | No. of engineering/design workflows supported by AI; % reduction in prototyping time. | No. of government services integrated with AI; % of citizens using AI-enabled e-services. | No. of national AI pilot projects across key sectors (health, agriculture, education). | No. of departments embedding AI in teaching or research workflows; faculty adoption rate. |
| **3 – Mastery of Infrastructure & Data Control** | HPC compute hours dedicated to AI R&D; % reliance on internal vs. external vendors. | No. of government registries connected through X-Road; uptime/reliability of secure data exchange. | No. of continental data centres established; % of AU states adopting federated data governance frameworks. | No. of petaflops available at TACC/Frontera for AI; % of system-wide data governed by shared policies. |
| **4 – Custom Model Development & Innovation** | No. of proprietary AI models developed (e.g., predictive maintenance, safety models); R&D spend on AI innovation. | No. of bespoke AI tools commissioned (e.g., anomaly detection, legal AI); % of ministries with in-house model development. | No. of AU-funded sector-specific AI innovations; % of R&D budget allocated to AI innovation. | No. of interdisciplinary AI projects funded; No. of custom AI curricula/programs (e.g., dual M.D./M.S.). |
| **5 – Advanced Deployment & Utilisation** | % of AI models deployed in production; No. of safety incidents prevented by AI; regulatory compliance metrics. | % of government services delivered via AI (*Bürokratt* adoption rate); citizen satisfaction ratings. | No. of AU AI ethical/legal frameworks ratified by member states; % of AI initiatives undergoing bias/impact audits. | % of academic programs integrating AI oversight policies; No. of clinical or administrative AI deployments. |
| **6 – Scaling & Reliability** | Global % of production processes AI-enabled; AI system uptime (SLA adherence). | No. of ministries adopting AI tools; % of national population regularly using AI-enabled services. | % of AU states with AI infrastructure rollouts; No. of continental workforce AI training initiatives. | Total enrolment in new AI-focused colleges/programs; % of UT campuses with federated AI governance structures. |

| Level | GM (Industry) | Estonia (National Government) | AU AI Strategy (Continental Policy) | UT System (Higher Education) |
|---|---|---|---|---|
| **7 – Ecosystem Collaboration & Societal Impact** | Participation in global AI/autonomous driving standards; No. of cross-industry consortia joined. | No. of open-source AI tools/platforms released (e.g., *Bürokratt* adoption abroad); ranking in global e-governance indices. | No. of AU contributions to global AI governance fora; % of cross-border AI collaborations launched. | No. of open datasets, models, or research outputs shared globally; citations/impact of UT AI research. |

**Table 14: Comparative Alignment of HE-AI Model with Existing AI Maturity Frameworks**

| HE-AI Level | Distinctive Contribution of HE-AI | PwC | Deloitte | Gartner | Enterprise AI Maturity Model |
|---|---|---|---|---|---|
| 0 – Initial Exposure & Curiosity | Recognises curiosity-driven, informal entry points as valid starting stages; emphasises inclusion and accessibility. | – | – | Awareness assumed | Stage 1 assumes workforce preparation already underway |
| 1 – Awareness & Orientation | Highlights trust formation and responsible-use guidelines alongside bounded task automation. | Awareness stage (assumed) | Starters (basic awareness) | Awareness stage | Stage 1 focuses on policy and education, but not trust-formation |
| 2 – Guided Application | Frames adoption as socially reinforced and collaborative; embeds literacy and training as motivations. | Active → Operational | Explorers | Pilots/active adoption | Stage 2 emphasises pilots, less on collaborative literacy |
| 3 – Autonomous Utilisation | Emphasises autonomy, sovereignty, and federation of infrastructure, not just technical readiness. | Operational | Pathseekers | Operationalisation | Stage 2–3 cover infrastructure; sovereignty/federation absent |
| 4 – Creation & Innovation | Innovation framed as creative, contextual, and ethically anchored, beyond ROI or competitiveness. | Systematised | Pathseekers/ Transformers | Early transformation | Stage 3 emphasises embedding processes, less on ethics/creativity |
| 5 – Responsible Deployment | Introduces an "ethics gate" where governance, lifecycle monitoring, and oversight are mandatory for scaling. | Transformational | Transformers | Transformation | Stage 3 embeds governance but no explicit "gate" before scaling |
| 6 – Human–AI Co-Evolution | Defines maturity as durable, federated, and adaptive scaling; stresses resilience and interoperability. | Transformational | Enterprise-wide embedding | Pervasive adoption | Stage 4 emphasises scale and integration, less on federation and resilience |
| 7 – Societal & Global Integration | Extends maturity beyond organisational advantage to stewardship, reciprocity, global interoperability, and societal good. | Not covered | Not covered | Not covered | Stage 4 focuses on enterprise future-readiness, not societal integration |